\documentclass[amsmath,amssymb,aps,prb,10pt,superscriptaddress,english,twocolumn,longbibliography]{revtex4-2}
\usepackage{amsmath,amssymb,amsfonts}
\usepackage{graphicx,epsfig}
\usepackage{dcolumn}
\usepackage{color,epstopdf}
\usepackage{tikz}
\usepackage{float}
\usepackage{bm}
\usepackage{latexsym}
\usepackage[normalem]{ulem}
\usepackage{hyperref}
\hypersetup{colorlinks=true,bookmarks=false,citecolor=blue,linkcolor=blue}
\usepackage{multirow}
\usepackage{braket}
\usepackage{xcolor}
\usepackage{color}
\usepackage[smalltableaux]{ytableau}
\usepackage[normalem]{ulem}

\begin{document}

\title{Ergodic and non-ergodic properties of disordered SU($3$) chains}

\author{Bhupen Dabholkar}
\email{dabholkhar@irsamc.ups-tlse.fr}
\affiliation{Laboratoire de Physique Th\'{e}orique, Universit\'{e} de Toulouse, CNRS, UPS, France}

\author{Fabien Alet}
\email{fabien.alet@cnrs.fr}
\affiliation{Laboratoire de Physique Th\'{e}orique, Universit\'{e} de Toulouse, CNRS, UPS, France}

\date{\today}

\begin{abstract}
Non-abelian symmetries are thought to be incompatible with many-body localization, but have been argued to produce in certain disordered systems a broad non-ergodic regime distinct from many-body localization. In this context, we present a numerical study of properties of highly-excited eigenstates of disordered chains with SU($3$) symmetry. We find that while weakly disordered systems rapidly thermalize, strongly-disordered systems indeed exhibit non-thermal signatures over a large range of system sizes, similar to the one found in previously studied SU($2$) systems. Our analysis is based on the spectral, entanglement, and thermalization properties of eigenstates obtained through large-scale exact diagonalization exploiting the full SU($3$) symmetry.
\end{abstract}
\maketitle
\section{Introduction}

Localization induced by the presence of strong disorder is one way for interacting quantum many-body systems to escape the prevailing thermalization paradigm epitomized by the eigenstate thermalization hypothesis (ETH)~\cite{Deutsch91,Srednicki94,Rigol08}. The resulting many-body localization (MBL)~\cite{Basko06,Gornyi05} has been the subject of many theoretical and experimental investigations (see reviews~\cite{Alet18,Abanin19}) and is the arena for a plethora of non-ergodic phenomena: absence of transport even at ``infinite temperature"~\cite{Basko06,Oganesyan07}, entanglement area law for highly-excited eigenstates~\cite{Bauer13,Serbyn13,Huse14}, memory of the initial state~\cite{Ros17}, logarithmic growth of entanglement after a quench~\cite{Znidaric08,Bardarson12,Serbyn13}, to mention a few highlights. These phenomena, which are best observed in numerical simulations of finite one-dimensional many-body systems at strong disorder, are ascribed to the presence of an extensive set of local integral of motions~\cite{Huse14,Serbyn13b,Ros15,Imbrie16} in the MBL phase. 
Currently there is a debate on whether MBL really exists in the thermodynamic limit even in one dimension, and/or whether the phase transition from an ergodic to a MBL phase as driven by disorder can be probed by finite-size numerics, as well as on how to interpret them~\cite{Sels21,suntajs20,sierant22,Morningstar22,Crowley22,Sels22,Bera17,Weiner19,Sels23,Abanin21,Sierant20,Panda20,Doggen18,Long23}. Independent of this, numerical simulations are relevant to describe the various experiments~\cite{Schreiber15,Choi16,Smith16,Luschen17,Roushan17,Xu18,Lukin19,Chiaro19,Rispoli19} probing MBL physics which are performed on finite length and time scales. 

The influence of symmetries on the existence of many-body localization is also very intriguing~\cite{Parameswaran18}. While MBL is expected (and observed in several numerical studies~\cite{Oganesyan07,Pal10,Luitz15,Sierant18,Mace19,Kjall14,Sahay21,Moudgalya20,Laflorencie22,Chandran14,Bahri15,Kuno19}) to be compatible with abelian symmetries, symmetry-enhanced topological order~\cite{Potter16,Vasseur16} and non-abelian symmetries~\cite{Potter16} have been argued not to be compatible with MBL as characterized by a complete set of local integral of motions. The key argument for non-abelian symmetries is that they would allow and protect degenerate localized eigenstates (with a degeneracy possibly exponentially large with system size) at finite energy density, which would be unstable to interactions: loosely speaking, degeneracies offer too many thermalization channels for creation of resonances. As a result, the system must either thermalize (this is the expectation for systems with continuous non-abelian symmetry~\cite{Protopopov17}) or maintain MBL while spontaneously breaking the non-abelian symmetry to an abelian subgroup, the latter scenario having been demonstrated in numerical studies of spin chains with non-abelian symmetries with finite-dimensional irreducible representations~\cite{Prakash17,Friedman18}. A further possibility theoretically remains, that some form of localization persists but not in the form of strict MBL with local integral of motions: in this category quantum critical glasses~\cite{Vosk13,Vasseur15} (which criticality reflects in eigenstates with logarithmic scaling of entanglement in one dimension) have been the early contenders to be described, even though most candidate models have been recently shown to present signs of instability towards thermalization in the thermodynamic limit~\cite{Ware21}. 

The thermalization process of systems enjoying a continuous non-abelian symmetry may however be non trivial. Ref.~\cite{Protopopov20} provides a detailed analysis of the simplest lattice model with SU($2$) symmetry, the random-bond Heisenberg spin-1/2 chain, where a broad non-ergodic regime is identified at finite size for strong enough disorder, which is distinct from conventional MBL (i.e. with local integrals of motions). This regime, identified within a real-space renormalization group (RSRG) approach \cite{Vosk13,Vosk14,Pekker14,Vasseur15}, is characterized by the absence of resonances on a certain length scale, and share the common feature of logarithmic scaling of entanglement with quantum critical glasses. This non-ergodic regime disappears for large enough system sizes $L>L_{\mathrm {erg}}^{\mathrm{RSRG}}(\alpha)$ where the ergodic length scale $L_{\mathrm{erg}}^{\mathrm{RSRG}}(\alpha)$ depends on disorder strength ($1/\alpha$) and can be estimated from RSRG, whose breakdown is driven by proliferation of multi-spin resonances~\cite{Protopopov17,Protopopov20}. These multi-spin resonances (contrasting with two-body resonances expected to cause the breakdown of standard MBL systems without non-abelian symmetries) have also been shown, on the basis of a similar RSRG approach, to be the source of instability towards thermalization for a large class of non-abelian symmetries~\cite{Ware21}.

Most of this analysis is performed using heuristic arguments or within the RSRG which is based on strong disorder and does not capture the ergodic phase. In view of experimental interest for non-ergodic matter, with some experimental platforms hosting states with non-abelian symmetries~\cite{Schreiber15,Luschen17,Gorshkov10,Cazalilla14}, it is important to test the validity of such (analytical) predictions with unbiased numerics on microscopic disordered Hamiltonians with non-abelian symmetries. 

The RSRG results of Ref.~\cite{Protopopov20} are accompanied by exact diagonalization (ED) results on highly-excited eigenstates of random SU(2) spin chains, and confirm the clear existence of a thermal regime at small enough disorder as well as of a non-ergodic regime below a length scale $L^*(\alpha)$. The determination of $L^*(\alpha)$ is not unique, but one intriguing way is through the observation of non-monotonous (with system size) spectral statistics, $L^*(\alpha)$ being the scale at which statistics change behavior. The length scales $L^*(\alpha)$ (obtained from ED on relatively small system sizes) and $L_{\rm erg}^{\rm RSRG}(\alpha)$ (obtained from RSRG on much larger systems) do not coincide, but it is expected that they are proportional, providing an indirect evidence for the non-trivial thermalization scenario of continuous non-abelian symmetric systems from statistics on exact eigenstates of random spin chains of moderate sizes. A further evidence for the existence of a non-ergodic regime comes from another recent computation~\cite{Saraidaris} of the dynamical behavior of SU($2$) and SU($3$) systems after a quench. Finally we also note that results on disordered Floquet systems with SU($2$) symmetry~\cite{Yang20} as well as on the Hubbard model (which enjoys SU($2$) symmetry) with a symmetry-preserving disorder~\cite{Prelovsek16,Zakrzewski18,Mierzejewski18,Protopopov19,Leipner19,Thomson23} are also consistent with the absence of full MBL, and indications of non-thermalization at strong disorder.

So far and to the best of our knowledge, thermalization properties of eigenstates with a continous non-abelian symmetry have been studied only in the SU($2$) case~\cite{Protopopov20,Yang20}. 
In this work, we present large-scale diagonalization results for SU($3$) chains with random bond couplings where we consider properties of eigenstates in the middle of the spectrum and probe for their ergodic/non-ergodic properties. We find evidence for a similar scenario as in the SU($2$) case where thermalization occurs for a sufficiently large length scale $L^{*,\mathrm{SU(}3\mathrm{)}}(\alpha)$, even though for ultimately strong disorder values, the system sizes we can reach do not allow to probe this length scale, and we have to rely on similarity of finite-size trends to the SU($2$) case or to smaller disorder. Technically, the SU($3$) case is more challenging, as naively the matrix sizes for chains of length $L$ scale as $3^L$ (instead of $2^L$). To mitigate this, we make a full usage of SU($3$) symmetry by performing diagonalization in each irreducible representation. This allows to reach chains of sizes up to $L=21$, instead of at best $L=15$ with standard computations in the usual U($1$) ($S^z$) basis. 

The plan of the paper is as follows. In Sec.~\ref{sec:model}, we introduce the disordered SU($3$) model, and the method that we use to obtain its eigenstates as well as details of the numerical computations (including the finite-size samples that we employ). Sec.~\ref{sec:results} contains our numerical results on spectral properties (Sec.~\ref{sec:gapratio}), distribution of expectation values of local observables (Sec.~\ref{sec:O}) as well as entanglement properties (Sec.~\ref{sec:EE}) of eigenstates. Our detailed analysis indicate clearly the existence of a thermal phase at weak disorder, and a broad non-ergodic regime at strong disorder, within the system sizes accessible to us. Finally, Sec.~\ref{sec:conc} offers a discussion on the ergodic length scale in our model in comparison with the results of Ref.~\cite{Protopopov20}, as well as computational perspectives on other problems of thermalization of systems with non-abelian symmetries. 

\section{Model and methods}
\label{sec:model}
The model considered in this work is a generalization of the Heisenberg spin-1/2 chain $H_{\rm SU(2)}=\sum_{i=1}^{L}J_{i} \bf{S}_{i} \cdot \bf{S}_{i+1} $, which enjoys SU($2$) symmetry, to SU($3$) symmetry where now the spin operators are replaced by permutation operators $P_{i,j}$:
\begin{equation}
    H_{\mathrm{SU(}3\mathrm{)}}=\sum_{i=1}^{L}J_{i}P_{i,i+1}.
    \label{eq:H}
\end{equation}
To each site we assign the fundamental representation of SU($3$) (single-box Young tableau) with a 3-dimensional local Hilbert space and a local basis $|c_i \rangle$ where $c=1,2,3$ is the ``color'' at site $i$. $P_{i,i+1}$ permutes colors at sites $i$ and $i+1$: $P_{i,i+1} | \dots c_i c'_{i+1} \dots \rangle =  | \dots c'_i c_{i+1} \dots \rangle $. 

The absolute value of the coupling constants $J_{i}$ are random variables drawn from the probability distribution
\begin{equation}
    P(|J|) = \frac{\alpha\Theta(1-|J|)}{|J|^{1-\alpha}}
\end{equation}
with $\alpha$ denoting the inverse strength of the disorder (the system is more disordered as $\alpha \to 0$). Furthermore, we choose the sign of $J_i$ to be positive/negative with an equal probability $1/2$, without loss of generality.

\begin{table*}
\begin{tabular}{|c|c|c|c|c|c|c|}
  \hline 
  $L$ & $15$ & $16$ & $17$ & $18$  & $19$ & $21$ \\
  \hline
  Young tableau & $(5,5,5)=\ydiagram{5,5,5}=\bullet^5$  & $(6,5,5)=\bullet^5 \ydiagram{1}$ & $(7,5,5)=\bullet^5 \ydiagram{2}$ & $(6,6,6)=\bullet^6$ & $(7,6,6)=\bullet^6 \ydiagram{1}$ & $(7,7,7)=\bullet ^7$  \\
  \hline
  Degeneracy  & 1 & 3 & 6 & 1 & 3 & 1  \\ \hline
  Hilbert space size $|{\cal H}|$ & 6006 & 36,036  & 136,136 & 87,516 & 554,268 & 1,385,670 \\ \hline 
\end{tabular}
  \caption{\label{tab:Hilbert}Chain samples used in this work, with the number of sites $L$, the shape of the Young-tableau corresponding to the irrep of SU($3$) (we use the simplified notation where $\bullet$ stands for singlets and $\bullet^n$ represent a $(n,n,n)$ (sub-)tableau), the degeneracy of this irrep in the spectrum of $H_{\mathrm{SU(} 3 \mathrm{)} }$, and the size of the Hilbert space in this sector.}
\end{table*}

The Hamiltonian $H_{\rm SU(3)}$ can be block-diagonalized in the irreducible representations (irreps) of SU($3$), each of which can be assigned a standard Young tableau with at most 3 lines. We use the orthogonal unit representation introduced in Ref.~\onlinecite{Nataf14} to be able to work directly in specific irreducible representations. Most calculations are carried out for the singlet irrep with rectangular Young tableaus with 3 lines. Since singlet tableaus restrict computations to chains of sizes $L$ multiple of 3, we also present some calculations for non-singlet tableaus. Table~\ref{tab:Hilbert} provides a description of the chain samples used in this work, with the number of sites $L$, the shape of the Young-tableau corresponding to the irrep of SU($3$), and the size of the corresponding Hilbert space in this irrep.

We use shift-invert diagonalization~\cite{Pietracaprina18} to obtain eigenstates of $H_{\rm SU(3)}$ in the middle of the spectrum ($\epsilon=0.5$ in standard notations where $\epsilon=(E-E_{\rm min})/(E_{\rm max}-E_{\rm min})$). The orthogonal unit representation of Ref.~\onlinecite{Nataf14} is particularly convenient for permutations operators $P_{i,i+1}$ as one obtains a sparse matrix representation for the Hamiltonian Eq.~\ref{eq:H} when open boundary conditions are used. This allows us to use efficient sparse linear algebra techniques~\cite{Roman:2023:ISR,strumpack,mumps} in the shift-invert method~\cite{Pietracaprina18}. For periodic boundary conditions, the matrices are less sparse (with roughly twice as many non-zero matrix elements), which does not allow to reach large sizes. We can simulate open chains with up to $L=21$ sites (with Hilbert space size of $1,385,670$), see Table~\ref{tab:Hilbert} for a list of all systems considered. For each irrep considered and for each disorder strength $\alpha$, we use at least 1000 disorder realizations, except for the $L=21$ singlet tableau where we use about $300$ disorder realizations for each $\alpha$ (due to the considerable simulation time needed for this large system). For each disorder realization, we collect between $50$ and $100$ eigenstates near $\epsilon=0.5$, which we refer to as mid-spectrum eigenstates below.

\section{Results}
\label{sec:results}

We consider three  different indicators to detect ergodic or non-ergodic behavior in eigenstates: spectral statistics in the form of the gap ratio (Sec.~\ref{sec:gapratio}), statistics of a local observable, the permutation operator $P_{i,i+1}$ between consecutive sites (Sec.~\ref{sec:O}) and the entanglement entropy (for different block sizes) for singlet states (Sec.~\ref{sec:EE}).

\subsection{Spectral Statistics}
\label{sec:gapratio}

\begin{figure}[h]
    \centering
    \includegraphics[width=0.99\columnwidth]{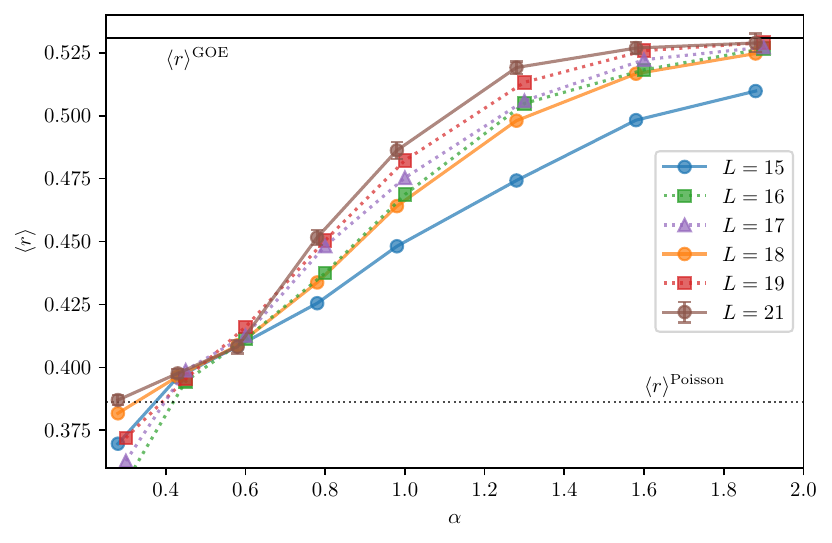}
    \includegraphics[width=0.99\columnwidth]{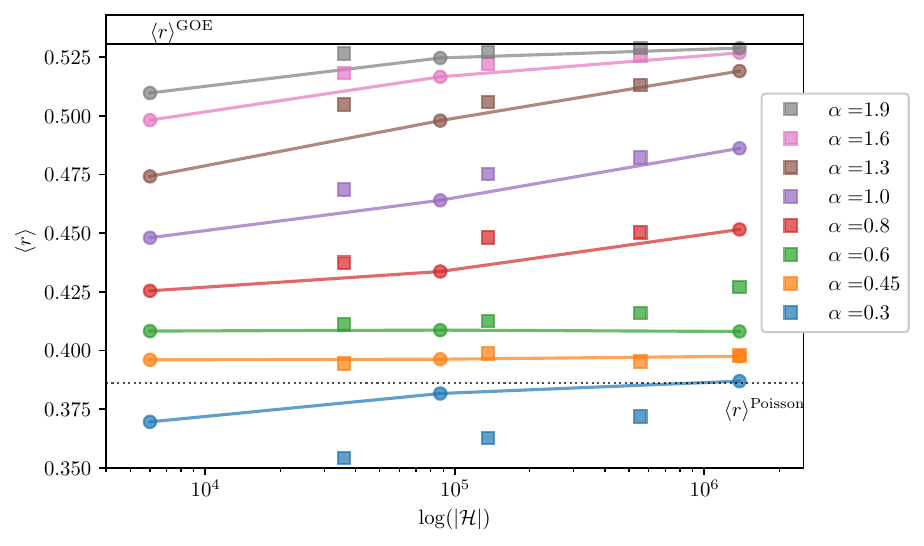}
    \caption{Top: Average Gap ratio as a function of (inverse) disorder strength $\alpha \in \{0,3. 0.45, 0.6, 0.8, 1, 1.3, 1.6, 1.9 \}$ for different chain lengths $L$. Solid lines represent data for singlet tableaus (chains of size $L=15,18,21$) and dashed lines are for non-rectangular tableaus (see Table ~\ref{tab:Hilbert}). Data for singlet tableaus for the values of $\alpha$ mentioned above have been slightly shifted to the left $(\alpha \rightarrow \alpha-0.02)$ for readability purposes. Error bars are smaller than symbol size, except for $L=21$ where they are explicitly given. The two limiting values $\langle r \rangle^\mathrm{GOE}$ and $\langle r \rangle^\mathrm{Poisson}$ for respectively thermal and Poisson statistics are also given. Bottom: Similar data, but as a function of the size of the Hilbert space $|{\cal H}|$ (in logarithmic scale) for different disorder parameters $\alpha$. Statistics for eigenstates in the singlet sector are joined by a solid line.}
    \label{fig:gapratio}
\end{figure}

We start our analysis with the consecutive gap ratio~\cite{Oganesyan07}
\begin{equation}
    r = \frac{\min(\Delta_{n},\Delta_{n+1})}{\max(\Delta_{n},\Delta_{n+1})}
\end{equation}
where $\Delta_n=E_{n+1}-E_n$ is the gap between the $n$-th and $(n+1)$-th energy levels $E_{n}$ and $E_{n+1}$, which is by now a standard measure of spectral statistics and thermal behavior for many-body systems. For thermal systems, the distribution $P(r)$ of the gap ratio  $0\leq r \leq 1$ is predicted for the Gaussian orthogonal ensemble (GOE, relevant for the real Hamiltonian Eq.~\ref{eq:H}) by random matrix theory~\cite{Atas2013} to be $P^\mathrm{GOE}(r)=\frac{27 (r+r^2)}{4(1+r+r^2)^{5/2}}$ with mean value $\langle r \rangle^\mathrm{GOE}\simeq 0.5307$. For integrable / MBL systems, a Poisson distribution $P^\mathrm{Poisson}(r)=2/(1+r)^2$ is expected, with mean value $\langle r \rangle^\mathrm{Poisson}=2 \ln(2)-1 \simeq 0.386$.

Finite-size dependence for the mean gap ratio (see Fig.~\ref{fig:gapratio}) clearly indicates that for large enough $\alpha\geq 0.8$, the system will become thermal as system size increases. For the largest size $L=21$, we can already observe statistics very close to the GOE prediction for $\alpha \geq 1.6$. For the disorder value $\alpha=0.45, 0.6$, the singlet tableaus strikingly show almost no system size dependence (see bottom panel of Fig.~\ref{fig:gapratio}) with a mean value between the Poisson and thermal values (the non-singlet tableaus show a slight finite-size tendency towards thermal behavior for $\alpha=0.6$). Finally, the behavior for the strongest disorder $\alpha=0.3$ is marked by unusually low values, even below the Poisson limit, except for the largest $L=21$ system. This behavior was also found in the SU($2$) case~\cite{Protopopov20} for the same value of $\alpha$ and is attributed to cutting-bond effects (small $J$ values tend to effectively cut the chain in two, resulting in almost degeneracies and low values for the gap ratio). This effect should disappear as a function of system size, which is what we indeed observe. 

Similar data in the SU($2$) case~\cite{Protopopov20} display a remarkable non-monotonous behavior (a decrease with small system size, followed by an increase on larger systems) for $\langle r \rangle$ for two intermediate values of disorder $\alpha=0.8, 1.0$, with a minimum for chain sizes around $L \in [14,18]$ (with corresponding Hilbert space sizes between $429$ and $4862$). We do not observe this behavior, except possibly for $\alpha=0.6$. We have checked (data not shown) that it does not occur specifically for systems of lower size than the smallest size $L=15$ (with Hilbert space size $6006$) presented in Fig.~\ref{fig:gapratio}, even though we are limited by the fact that there are not many possibilities available for this range of Hilbert space sizes for SU($3$) systems. It is also possible that this non-monotonous effect could arise at a larger length scale $L>21$ for $\alpha=0.45$.

Let us finally mention that open boundary conditions typically produce lower values of mean gap ratio than periodic boundary conditions (this should be a $1/L$ effect, albeit non-negligible for the moderate system sizes that we can probe). This was observed in other systems~\cite{Chanda20}, and we could also check this behavior for $L=12,15$ (data not shown).

\begin{figure}[h]
    \centering
    \includegraphics[width=0.99\columnwidth]{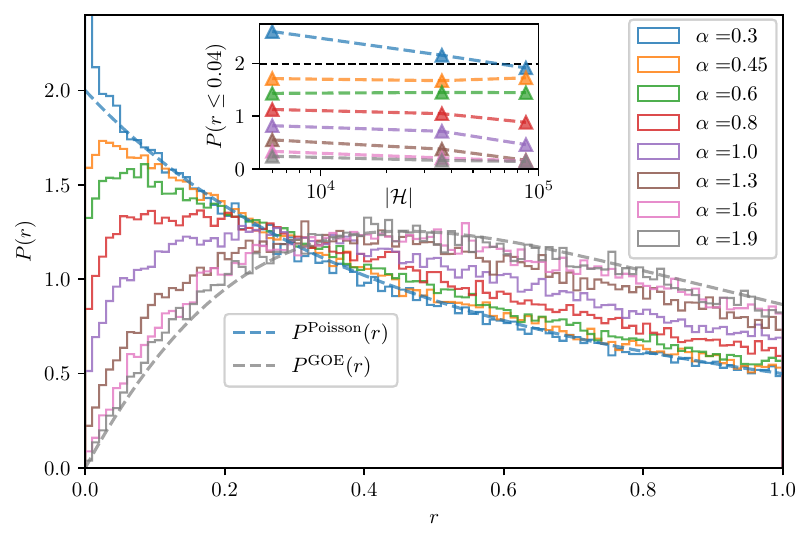}
    
    \caption{Probability distribution of gap ratio $P(r)$ for different disorder strengths for a $L=18$ chain. The dashed lines present the reference Poisson and GOE (thermal) dsitributions for comparison. Inset: Hilbert space size dependence of the level repulsion effect $P(0)$ (as approximated by $P(r \leq 0.04)$) for the same values of disorder.}
    \label{fig:Pr}
\end{figure}

Fig.~\ref{fig:Pr} offers a comparison of the probability distribution $P(r)$ for $L=18$ singlet eigenstates for the various strengths of disorder $\alpha$, to the limiting cases of the thermal distribution $P^\mathrm{GOE}(r)$ and Poisson distribution $P^\mathrm{Poisson}(r)$. For the strongest disorder $\alpha=0.3$, the distribution is very close to Poissonian except for low values of $r$ where it exhibits a sub-Poissonian value $P(0)>2$, corresponding to effective level attraction and low values of mean gap ratio, which can again be explaining by the cut-chain effects. For larger disorder $\alpha=0.45,0.6$, the distributions display a maximum for finite small values of $r$, and are thus markedly different from Poisson, even though their large $r$ behavior then closely joins the Poisson tail. For larger values of $\alpha$, a crossover is observed towards the thermal distribution which is reasonably well reproduced for $\alpha=1.6$ and $\alpha=1.9$. 

In the inset of Fig.~\ref{fig:Pr}, we analyze the level repulsion effect in more detail, by considering the finite-size dependence of $P(r\leq 0.04)$ (as a proxy to $P(0)$) for all values of $\alpha$ for singlet tableaus ($L=15,18,21$). The dependence is very similar to the one of mean gap ratio statistics: for $\alpha \geq 0.8$, there is clear evidence of level repulsion in the thermodynamic limit (as $P(0)$ decreases) whereas for $\alpha=0.45,0.6$, $P(0)$ seems approximately constant within the range of studied sizes. Finally the cutting-chain effects at very strong disorder $\alpha=0.3$ manifest in values of $P(0)$ larger than the Poisson value $P(0)=2$, but this effect disappears as the system size is increased.

The conclusion of the gap ratio analysis is that systems with strong disorder ($\alpha\leq 0.6$) clearly manifest evidence of non-thermal behavior for the system sizes $L\leq 21$ studied, which cannot be attributed to very small values of bond coupling that will tend to artificially increase level repulsion, falsely mimicking Poisson statistics (this effect is observed for $\alpha =0.3$ and its finite-size dependence understood). Indeed the distributions of $P(r)$ in this non-thermal regime are not Poissonian, especially at low values of $r$. For $\alpha=0.45$, the broad extent of Hilbert space sizes (from $10^3$ to $10^6$) over which the spectral statistics do not show tendency to either thermalization or Poisson behavior is quite remarkable.

\subsection{Local observables}
\label{sec:O}

\begin{figure*}[t!]
    \centering
    \includegraphics[width=1.3\columnwidth]{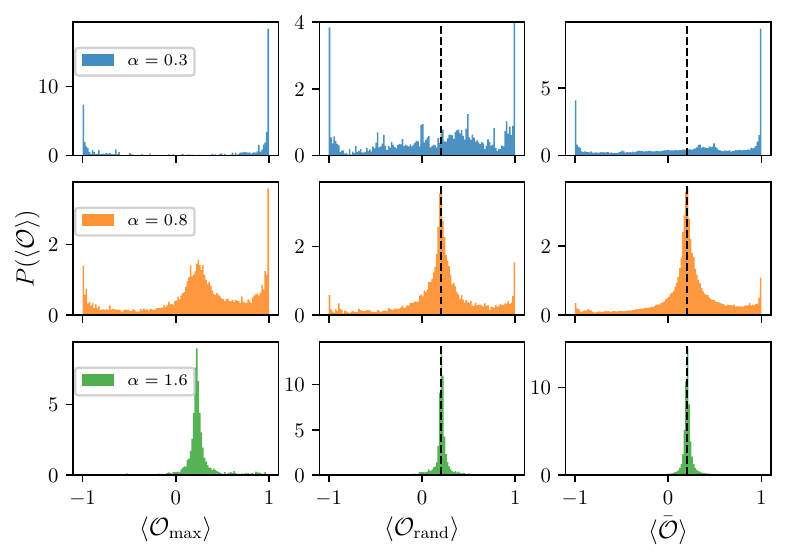}
    \caption{Distribution of the local observable ${O} = P_{i,i+1}$ shown for 3 disorder values $\alpha=0.3,0.8,1.6$ and for system size $L=21$. Left panel: $P(\langle {\cal O}_\mathrm{max} \rangle)$, for sites $i_{\rm max},i_{\rm max}+1$ with the strongest coupling in the chain; middle panel $P(\langle {\cal O}_\mathrm{rand} \rangle)$, for the site the further away $i=(i_{\rm max} + L/2) \mod L$; and right panel the average $P(\langle {\cal O}_\mathrm{rand} \rangle)$ over all bonds $i,i+1$. The dashed line corresponds to the expected thermal value $\langle {\cal O}_\mathrm{thermal} \rangle \simeq 0.2$ for $L=21$, as obtained from a random state, which is quite different from the expected thermodynamical value $1/3$. }
    \label{fig:PO}
\end{figure*}

We now proceed to results for the the distribution of a local observable $\braket{n|{\cal O}|n}$ with $\ket{n}$ mid-spectrum eigenstates. More specifically we consider the local permutation operator ${\cal O} = P_{i,i+1}$ between sites $i$ and $i+1$. The left panels of Fig.~\ref{fig:PO} show the distribution where $P_{i,i+1}$ is measured for $i=i_{\rm max}$ for the strongest coupling $J_{i,i+1}=J_{\rm max}$ in the chain, the middle panels for $i$ the further away from this strongest coupling $(i_{\rm max} + L/2) \mod L$ (which should correspond to a random coupling), and the right panels for the observable averaged over all possible values of $i,i+1$. The data are presented for three different representative disorder values $\alpha=0.3, 0.8$ and $1.6$ and for the largest system $L=21$.

In the limit of very large disorder $\alpha=0.3$, the distribution of $\langle {\cal O}_\mathrm{max} \rangle$ is dominated by two sharp peaks at ${\cal O}_\mathrm{max}=+1$ and ${\cal O}_\mathrm{max}=-1$, the latter being approximately half the size of the largest peak. This can be understood by adapting the argument put forward in the RSRG computation for the SU($2$) case: in the limit of infinite disorder, the strongest bond is decimated first, and the two fundamental representations of SU($3$) are coupled to form either a 2-box symmetric \ydiagram{2}~(with degeneracy $6$ and for which ${\cal O}=+1$) or antisymmetric \ydiagram{1,1}~(with degeneracy $3$ and ${\cal O}=-1$) irrep. This explains the two peaks at $\pm 1$ in the distribution $P({\cal O}_\mathrm{max})$ and their relative height. The opposite bond $(i_{\rm max} + L/2) \mod L$ would be decimated later in the decimation procedure, resulting in intermediate values with neverthelss the most likely values being again $\pm 1$. This is exactly what is observed in the top middle panel for $P({\cal O}_\mathrm{rand})$ at $\alpha=0.3$. Finally, the average over all possible bonds (right panel) corresponds to averaging over all levels of decimation, resulting in a distribution $P(\langle \bar{{\cal O}} \rangle)$ close to the one for the strongest bond $P(\langle {\cal O}_\mathrm{max} \rangle)$. We thus find that our data for $P(\langle {\cal O} \rangle)$  for the SU($3$) case are compatible with the strong disorder scenario advocated at small $\alpha$ in the SU($2$) case, namely that eigenstates can be considered (for moderate systems sizes) to be close to those obtained from a strong disorder RSRG procedure.

We now continue with the smallest disorder $\alpha=1.6$. We observe a close to Gaussian distribution around a most likely value ${\cal O}^*\simeq 0.2$, for all three distributions $P(\langle {\cal O}_\mathrm{max} \rangle )$ , $P(\langle {\cal O}_\mathrm{rand} \rangle )$ , $P(\langle \bar{\cal O} \rangle )$, which as we argue now matches the expectation of ETH. In the thermal regime and for mid-spectrum eigenstates, the observable should be the average taken uniformly over all the states formed by coupling two fundamental tableaus on  sites $i$ and $i+1$. It is easy to see that this average should be $1/3$, however for finite (small) systems this may not be exactly the case. Indeed by randomly sampling Hilbert space of finite systems, we find that the expected limit $1/3$ is reached only slowly as $1/L$~\footnote{More precisely, we find that the the expression $\langle \mathrm{Random}(L) | \bar{\cal O}| \mathrm{Random}(L) \rangle =\frac{1-8/(L-1)}{3}$ fits very well the data for singlet tableaus}. In particular, for the singlet size $L=21$ (singlet tableau), the random sampling provides an average $\langle \bar{\cal O} \rangle \simeq 0.200$, which is matching very well with the ``ergodic" peaks at ${\cal O}^*\simeq 0.2$ in Fig.~\ref{fig:PO} for $\alpha=1.6$.

The situation for the intermediate disorder $\alpha=0.8$ is very instructive. For random $P(\langle {\cal O}_\mathrm{rand} \rangle)$ and average $P(\langle \bar{\cal O} \rangle)$ distributions, the ergodic peak at ${\cal O}^*\simeq 0.2$ dominates but a non-zero background for all other possibles values of ${\cal O}$ is present with peaks at $\pm 1$. We interpret this as a tendency to ETH for this observable for most eigenstates, with remnants of non-thermal behavior. For the strongest coupling, the distribution $P(\langle {\cal O}_{\rm max} \rangle)$ is close to trimodal with a central peak at ${\cal O}^*\simeq 0.2$, but slightly more dominant sharp peaks at $\pm 1$, together with the same background of non-zero probability for other values of $\langle {\cal O}_{\rm max} \rangle$. This behavior is also compatible with a tendency to thermalization for this value of $\alpha$ but with a slower convergence / stronger finite-size effect for this specific observable, which indeed is atypical as associated to the strongest bond in the Hamiltonian.

\begin{figure}[t]
    \centering
    \includegraphics[width=0.99\columnwidth]{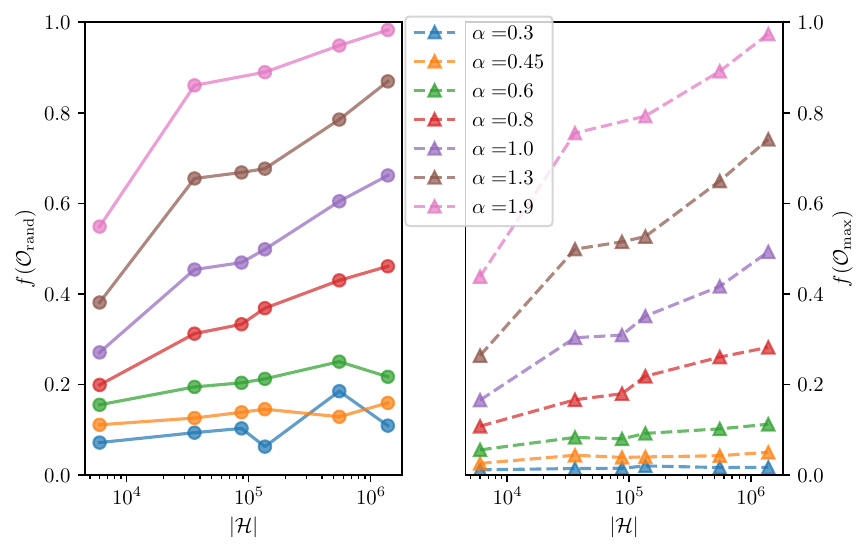}
    \caption{Fraction of eigenstates that belong to the central ergodic peak, with $\hat{\cal O}_{max}$ in straight lines and $\hat{\cal O}_{rand}$ with dashed lines. ``Belonging'' to the central peak actually means $O$ in $[O^*-\delta,O^*+\delta]$ with $\delta=1/4$.}
    \label{fig:fraction}
\end{figure}

In order to probe finite-size effects for these distributions, we consider the scaling (with the size $|{\cal H}|$ of Hilbert space) of the fraction $f({\cal O})$ of eigenstates which belong to the ``ergodic'' peak close to $~ 0.2$, similar to what was done in Ref.~\cite{Protopopov20}. We consider the fraction of states which have $\langle O \rangle$ in $[O^*(|{\cal H}|)-\delta,O^*(|{\cal H}|)+\delta]$, with $O^*(|{\cal H}|)$ the most probably value obtained from a random-state sampling and where $\delta=1/4$, as chosen to be able to make a comparison with Ref.~\cite{Protopopov20} which consider $\delta=1/8$ for an observable in the range $[-3/4,1/4]$. Our results in Fig.~\ref{fig:fraction} are compatible with a fraction $f$ that tends to $1$ as predicted by ETH, for all values of $\alpha \geq 0.6$, for both ${\cal O}_\mathrm{max}$ and ${\cal O}_\mathrm{rand}$, albeit with different speeds of convergence. For the two strongest disorder strenghts, we find an overall slight tendency for this fraction to increase with Hilbert space size, albeit obscured by a finite-size and Young-tableau shape dependence. It is clear nevertheless that on these small system sizes, the ETH prediction $f \rightarrow 1 $ is far from being reached since $f<0.2$ for $\alpha \leq 0.6$, meaning that the eigenstates are clearly non-thermal for these ranges of $L$ and disorder $\alpha$. This is overall in agreement with the existence of a finite-size regime with non-thermal behavior for this SU($3$) disordered system. Comparing to SU($2)$ disordered chains with the same Hilbert space sizes (see Fig.~9 in Ref.~\cite{Protopopov20}), we find similar ergodic fractions for $\alpha=0.6, 0.8$, and SU($2$) systems to be very slightly more thermal than SU($3$) systems for $\alpha=1.0$. 

\subsection{Entanglement entropy}
\label{sec:EE}

We consider a bipartition of the chain in two parts $A$ and $B$ constituted respectively by the $L_A$ first sites $A=\{ i=1 \dots L_A\}$ and $L_B$ remaining ones $B=\{ i=L_A \dots L\}$ with $L_A+L_B=L$. Defining the reduced density matrix $\rho_A = {\rm Tr}_B | n \rangle \langle n |$ of an eigenstate $|n \rangle$ for such a bipartition, the entanglement entropy of this eigenstate is given by:
$$
S(|n\rangle)=-{\rm Tr}_A \rho_A \log (\rho_A).
$$

\begin{figure}[t]
    \centering
    \includegraphics[width=0.99\columnwidth]{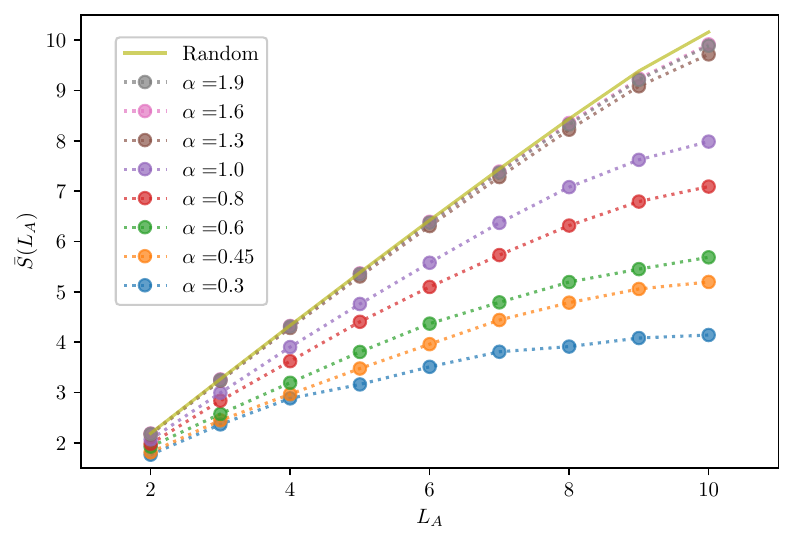}
    \includegraphics[width=0.99\columnwidth]{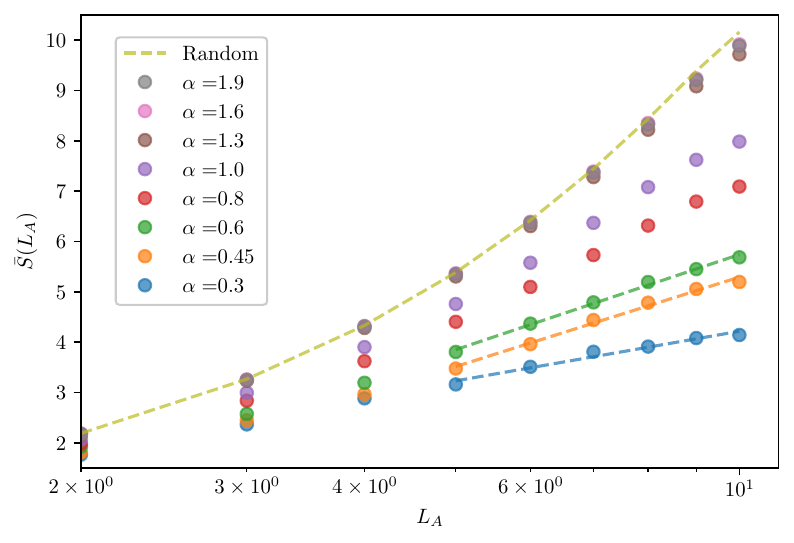}
    \caption{Both panels represent entanglement entropy $S$ of mid-spectrum eigenstates versus block size $L_A$, as a function of disorder parameter $\alpha$ and for a chain with $L=21$ sites. The solid line is the result for a wave-function on a similar $L=21$ size with Gaussian-distributed coefficients, corresponding to the Page~\cite{Page93} entanglement volume law. Top panel: linear scale for the block size $L_A$, where dashed lines are guides to the eye. Bottom panel: Logarithmic scale for $L_A$. The dashed lines are fits to a logarithmic scaling $S(L_A)=a+b \log (L_A)$ for the three strongest disorders and $L_A \in [5,10] $.}
    \label{fig:EE}
\end{figure}

We present results of entanglement entropy (following Ref.~\cite{NatafDMRG} for its computation in the orthogonal unit basis) for the non-degenerate singlet mid-spectrum eigenstates of the largest $L=21$ system we could simulate, as a function of block size $L_A$ in Fig.~\ref{fig:EE}. The results are symmetric with respect to inversion of $L_A$ and $L_B=L-L_A$, thus we only show data up to $L_A=10$. For small disorder (large $\alpha \geq 1.3$), the entanglement entropy follows a very clear volume law scaling (with $S$ growing linearly with the block size, see top panel of Fig.~\ref{fig:EE}), with a coefficient matching with the expected Page~\cite{Page93} behavior obtained by considering the entanglement entropy of random vectors (distributed on the Haar sphere), presented as a solid line in Fig. ~\ref{fig:EE}. For intermediate disorder $\alpha =0.8, 1.0$, the growth of entanglement appears linear for small $L_A$ but data bends as $L_A$ reaches $L/2$, due to the finite block-size limitation. For larger disorder (particularly for $\alpha=0.3, 0.45$), we observe a slow growth of entanglement, which we find compatible with a logarithmic growth, within the system sizes that can be addressed (see bottom panel of Fig.~\ref{fig:EE} where fits to logarithmic scaling of $S(|n \rangle)$ are shown). This contrasts not only with the volume law at large $\alpha$, but also to a strict area law expected for standard MBL eigenstates. This logarithmic ``subthermal" behavior has been predicted for the entanglement entropy in the intermediate, finite-size, regime of SU($2$) disordered chains by the RSRG calculations~\cite{Protopopov17,Protopopov20}, but was not strictly speaking observed in the finite-size exact diagonalization results of Ref.~\cite{Protopopov20}.

\section{Discussion and conclusions}
\label{sec:conc}

Through the lens of various indicators (spectral statistics, expectation values of local observables, entanglement entropy), we examined the behavior of mid-spectrum eigenstates for disordered chains with the non-abelian SU($3$) symmetry. While the data at weak disorder perfectly match the expectations for a thermal phase, our numerical analysis on finite sizes at strong disorder (small values of $\alpha$) points towards a non-thermal regime. 

Could our data be interpreted as signatures of a phase transition from a (standard) MBL phase (or regime) to a thermal phase at a strong value disorder (for, say,  $\alpha \simeq 0.45$), similar to the interpretation of similar data in the standard MBL model~\cite{Luitz15}? We believe this is not the case for the following reasons: first, there is no regime where the spectral statistics exhibit perfect Poisson statistics. The level repulsion at $P(0)$ is either sub-Poissonian (for $\alpha=0.3$) on the system sizes considered, or already non-Poissonian for the nearby disorder strength $\alpha=0.45$. Second, even at very large disorder $\alpha=0.3$, the distribution of expectation values of local observable has a non-zero background for some local observable (such as ${\cal O}_\mathrm{random}$ in top middle panel of Fig.~\ref{fig:PO}). Finally, the entanglement entropy displays a behavior compatible with a logarithmic growth (as a function of block size) in the region $\alpha=0.3-0.6$. While it is hard to distinguish a log from a small power-law, our data appear to rule out an area law that standard MBL eigenstates would follow. 

We thus interpret our data as completely compatible with the existence of a non-thermal, finite-size, regime in the region $\alpha \approx [0.3, 0.6]$ for systems below an ergodic length scale $L^{*, \mathrm{SU(} 3 \mathrm{)}}$, similar to the SU($2$) situation predicted by the RSRG computation~\cite{Protopopov17,Protopopov20}. It is difficult to conclude on whether the thermal length scale $L^*(\alpha)$ above which the system converges to the ETH is larger or not for SU(3) systems than SU(2).
The gap ratio data do not explicitly show a minimum behavior (except maybe for $\alpha=0.6)$ as in the SU($2$) case which would help identifying this length scale. This is possibly due to the lack of systems with intermediate Hilbert space sizes which could help visualizing this behavior. On the other hand, the very slightly more thermal data for the fraction of thermal eigenstates ($f$ in Sec.~\ref{sec:O}) for SU($2$) for fixed Hilbert space size can possibly and at best indicate that $L^{*, \mathrm{SU(}3 \mathrm{)} }(\alpha=1) \gtrsim L^{*,\mathrm{SU(}2\mathrm{)}}(\alpha=1) \ln(2) / \ln(3) $, which is not predictive. Overall the overall agreement of the evidence of non-thermal data for $\alpha=0.3, 0.45$,  and intermediate behaviour for $\alpha=0.6,0.8$ in a similar range of system sizes for SU($2$) (Ref.~\cite{Protopopov20}) and SU($3$) (our data) point towards the hypothesis that these two length scale are probably very similar $L^{*,\mathrm{SU(}3\mathrm{)}}(\alpha) \sim L^{*,\mathrm{SU(}2\mathrm{)}}(\alpha)$. It would be very interesting to have RSRG predictions for the SU($3$) case for this length scale, as well as in general to compare to our exact diagonalization data when possible. This would require extending~\cite{Vosk13,Vosk14,Pekker14,Vasseur15} to highly-excited eigenstates the RSRG procedure developped for ground-states of random SU($N$) symmetric systems~\cite{Hoyos04}.

Our results are in agreement with the general arguments forbidding a true MBL phase for systems with non-abelian symmetries~\cite{Potter16} as well as the recent dynamical study ~\cite{Saraidaris} on the same SU($3$) systems (performed on larger system sizes, up to $L=48$, but up to a finite maximal time $t_{\mathrm max}=500$) which also find a sub-thermal behavior for values $\alpha=0.3$ and $\alpha=0.5$. This dynamical analysis indicates a slightly more thermal behavior for SU($3$) then for SU($2$), for these values of $\alpha$ and lengths. 

Our work exploits entirely the SU($3$) symmetry by performing exact diagonalization in each irrep thanks to the use of the orthogonal unit basis introduced in Ref.~\cite{Nataf14}. This basis could be employed to systems with SU($4$) symmetry: for singlet eigenstates, system sizes $L=16$ and $L=20$ are reachable within shift-invert computations. Another interesting application of the orthogonal unit basis (which allows for diagonalization or time-evolution of larger systems than in the standard U($1$) basis) would be to test in detail recent predictions of a non-abelian version of the ETH~\cite{Murthy23,Majidy23}, which has been argued to present different finite-size convergence to the thermal ensemble than the standard ETH. 

\begin{acknowledgements}

We thank D. Abanin, N. Laflorencie and P. Nataf for fruitful discussions. This work benefited from the support of the joint PhD program between CNRS and IISER Pune, and of the Fondation Simone et Cino Del Duca. 
We acknowledge the use of HPC resources from CALMIP (grants 2022-P0677 and 2023-P0677) and GENCI (projects A0130500225 and A0150500225), as well as of the PETSc~\cite{petsc-user-ref,petsc-web-page,petsc-efficient}, SLEPc~\cite{Hernandez:2003:SSL,Hernandez:2005:SSF,Roman:2023:ISR}, MUMPS~\cite{mumps} and Strumpack~\cite{strumpack} sparse linear algebra libraries.

\end{acknowledgements}

\bibliography{su3}

\end{document}